\documentclass[prl,aps,twocolumn]{revtex4}

\usepackage{amssymb}
\usepackage{graphicx}
\usepackage{amsmath}

\begin{document}

\title{Magnetic Field Induced Coherence-Incoherence Crossover
in the Interlayer Conductivity of a Layered Organic Metal}
\author{M. V. Kartsovnik$^1$}
\author{P. D. Grigoriev$^2$}
\author{W. Biberacher$^1$}
\author{N. D. Kushch$^3$}

\affiliation{$^1$Walther-Meissner-Institut, Bayerische Akademie der
Wissenschaften, D-85748 Garching, Germany}
\affiliation{$^2$ L. D. Landau Institute for Theoretical
Physics, Russian Academy of Sciences, 142432 Chernogolovka, Russia}
\affiliation{$^3$ Institute of Problems of Chemical Physics,
Russian Academy of Sciences, 142432 Chernogolovka, Russia}

\begin{abstract}
The angle-dependent interlayer magnetoresistance of the
layered organic metal $\alpha$-(BEDT-TTF)$_2$KHg(SCN)$_4$ is
found to undergo a dramatic change from the classical conventional
behavior at low magnetic fields to an anomalous one at high fields.
This field-induced crossover and its dependence on the sample purity
and temperature imply the existence of two
parallel channels in the interlayer transport: a classical
Boltzmann conductivity $\sigma_{\mathrm{c}}$ and an
incoherent channel $\sigma_{\mathrm{i}}$. We propose
a simple model for $\sigma_{\mathrm{i}}$ explaining its metallic
temperature dependence and low sensitivity to the inplane
field component.
\end{abstract}
\maketitle

Dimensional crossovers and their
influence on transport properties and electronic states is
a long-standing and still controversial issue in the field
of highly anisotropic correlated conductors, such as
superconducting cuprates, cobaltates, organics, intercalated
compounds, etc.
One of the most frequently discussed mechanisms of breaking 
the interlayer band transport in a layered metal is due to
scattering. If the scattering rate $\tau^{-1}$ is larger than the
interlayer hopping rate, $\tau_h^{-1} \sim t_{\perp}/\hbar$, the
quasiparticle momentum and Fermi surface are only defined within
conducting layers, i.e. become strictly two-dimensional (2D).
Nevertheless, as long as the charge transfer between two adjacent
layers is determined by direct one electron tunneling
("weakly incoherent" regime \cite{kenz98}), the interlayer
resistivity $\rho_{\perp}(T)$ is predicted to be identical to that in
the fully coherent three-dimensional (3D) case \cite{kuma92,maslov}.
At increasing temperature, the conductivity due to direct tunneling
decreases and other conduction mechanisms associated,
e.g., with small polarons \cite{lund03,ho05} or resonant impurity tunneling
\cite{abri99,maslov} may come into play. This may lead to a crossover
from a low-temperature metallic to a high-temperature, seemingly,
nonmetallic temperature dependence of $\rho_{\perp}$.
Such a scenario is qualitatively consistent with a nonmonotonic
dependence $\rho_{\perp}(T)$ with a maximum at $T_m \sim 100$~K
reported for various layered materials 
\cite{lavr98,wang05,bura92,valla02,anal06}.
However, it does not explain the fact that the resistivity anisotropy
in many of these compounds grows continuously upon cooling deep into
the metalliclike regime of $\rho_{\perp}(T)$
\cite{lavr98,wang05,bura92,zver06}.

In addition to the latter apparent inconsistency, recent
magnetotransport experiments have revealed a low-temperature
behavior strongly violating theoretical predictions.
The interlayer resistance $R_{\perp}$ of a, presumably, weakly
incoherent sample of the organic metal
$\alpha$-(BEDT-TTF)$_2$KHg(SCN)$_4$ has been found to be
insensitive to a strong magnetic field applied parallel to layers
\cite{kart06}.
This is, in particular, reflected in a broad dip in the angular
dependence of magnetoresistance which is centered at $\theta=90^{\circ}$
and scales with $B_{\perp}=B\cos\theta$, where $\theta$ is the angle
between the
field and the normal to layers. While a similar dip in the
angle-dependent magnetoresistance (AMR) has been observed on
a number of other layered materials with different inplane
Fermi surface topologies \cite{dann95,kang03b,wosn02,kura03},
its origin remains unexplained.

For the organic conductor (TMTSF)$_2$PF$_6$ characterized by
a flat, weakly warped Fermi surface the anomalous dip structure
was reported for a field rotation in the plane of the Fermi sheets
\cite{dann95,chas98}. It was, however, noticed \cite{chas98} that
the dip  develops only at a high
enough magnetic field $B > 1$~T; at low fields the curves
$R_{\perp}(\theta)$ display a conventional shape with a maximum at
the field parallel and a minimum at the field perpendicular to layers.
The dramatic change of the AMR behavior was interpreted as a result
of a field-induced confinement of conducting electrons.
Semiclassically, the excursion
of a charge carrier across the layers is restricted by a strong 
inplane magnetic field $B_{\|}$ and limited to within one layer when 
$B_{\|}\geq B_c=4t_{\perp}/edv_F$, 
where $e$ is the elementary charge, $d$ is the interlayer period, and 
$v_F$ is the Fermi velocity.
This was suggested to lead to a dimensional crossover
and a consequent breakdown of the Fermi-liquid behavior
\cite{clarke1}.
While the field-induced confinement scenario \cite{clarke1}
describes qualitatively a number of features of the magnetoresistance
in (TMTSF)$_2$PF$_6$, it still does not provide a consistent explanation
for the dip around $\theta =90^{\circ}$. It remains also unclear why
the crossover field increased with temperature in the experiment
\cite{chas98}. Further, as it will be shown
below, the crossover between the low-field, conventional and high-field,
anomalous AMR can also be observed on a system possessing a cylindrical
Fermi surface. It is unclear, to what extent the field-induced confinement
can be effective in this case.

Here, we report on the crossover in the shape of the angle-dependent
interlayer magnetoresistance of $\alpha$-(BEDT-TTF)$_2$KHg(SCN)$_4$.
All the measurements were done under a pressure of $\approx 6$~kbar
in order to suppress the density-wave formation and stabilize the
normal metallic state \cite{kart07} with a well defined Fermi surface
consisting of a pair of open sheets and a cylinder \cite{mori90a}.
We show that the field-induced confinement model \cite{clarke1} is
inconsistent
with the evolution of the crossover with temperature and sample
purity. On the other hand, the observed behavior is strongly suggestive
of two parallel contributions to the interlayer conductivity: a classical
Boltzmann channel, $\sigma_c$, and an anomalous, incoherent channel,
$\sigma_i$. We propose a possible explanation of the field and temperature
dependence of $\sigma_i$, without invoking non-Fermi liquid effects.

Figure 1 shows AMR patterns from two samples of
$\alpha$-(BEDT-TTF)$_2$KHg(SCN)$_4$ recorded at $T=1.4$~K, at different
field intensities, $B= 0.12, 0.5, 3$, and 15~T. 
The transport current is
applied perpendicular to the $ac$ plane, that is the plane of conducting
layers. Azimuthal angle $\varphi$ measured between the field projection 
on the $ac$ plane and the $a$ axis, the latter being perpendicular to 
the open Fermi sheets, is $\approx 20^{\circ}$ for both samples. 
The oscillatory behavior, particularly
pronounced at high fields is due to angular magnetoresistance oscillation
and Shubnikov-de~Haas effects, as described in detail elsewhere
\cite{andr08}. It reveals a high crystal quality of both samples, 
providing an estimate for the transport scattering time,
$\tau_1\approx\tau_2/3 \sim 5$~ps for samples \# 1 and
\# 2, respectively. In the rest of the paper we focus on
the nonoscillating magnetoresistance component.
\begin{figure}
\includegraphics[width=.85\linewidth]{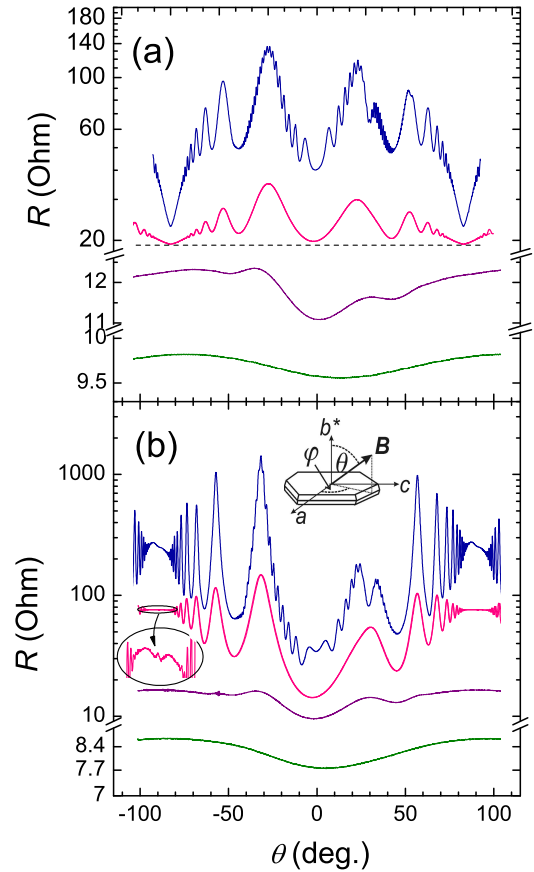}
\caption{(color online). Angle-dependent interlayer
magnetoresistance of a relatively dirty sample,
\# 1, of $\alpha$-(BEDT-TTF)$_2$KHg(SCN)$_4$ in the
high-pressure metallic state recorded at $T=1.4$ K, at
magnetic fields (bottom to top): 0.12, 0.5, 3, and 15 T;
$\varphi \approx 20^{\circ}$ (b) Same for
a very clean sample, \# 2. The upper inset illustrates the
definition of angles $\theta$ and $\varphi$; the lower inset:
enlarged fragment of the 3 T curve showing a small "coherence 
peak".}
\label{AMR}
\end{figure}

The low-field curves shown in Fig. 1 have a conventional form for both
samples: the AMR is maximum(minimum) at the field directed nearly
parallel(perpendicular) to layers. At fields above $\sim 1$~T, a broad
dip around $\theta =\pm 90^{\circ}$ develops in the AMR of sample \#1.
Already at $B=3$~T the magnetoresistance displays an absolute minimum
at the field aligned parallel to layers (and perpendicular to the current!).
This crossover in the AMR shape is very similar to that
reported for (TMTSF)$_2$PF$_6$ \cite{chas98}. Note, however, that in the
present case the Fermi surface contains, besides open sheets, a
cylindrical part and $\varphi \approx 20^{\circ}$ corresponds to the
field rotating close to the plane perpendicular to the open sheets.
Such a geometry is clearly unfavorable for the
field-induced confinement scenario \cite{clarke1}.

A comparison between the AMR of two samples shown in Fig. 1 reveals yet
another disagreement with the field-induced confinement model.
While the confinement field $B_c$ is formally independent of the 
scattering time, the model implies a sufficiently high $\tau$, so 
that the strong field criterion,
$\omega_c\tau > 1$ ($\omega_c$ is the
characteristic frequency of orbital motion in a magnetic field), 
is fulfilled. Therefore, the effect should be
seen, first of all, in clean samples.
By contrast, in our case the crossover is observed in the relatively
dirty sample \# 1, whereas the cleaner sample \# 2
preserves the normal anisotropy up to the highest field applied.

To explain the observed behavior, it is instructive to study the 
magnetoresistance of sample \# 1 as a function of magnetic 
field, aligned parallel to layers, and its evolution with
temperature. The relevant data is shown in Fig. 2 in the form of
a Kohler plot. Here, $R_0(T)$ is the zero-field resistance shown in
the inset and the normalized field-dependent interlayer
conductivity $\sigma(B,T)/\sigma(0,T)$ has been obtained from $R(B)$
measurements, taking into account that $\sigma(B) \propto 1/R(B)$
in our quasi-2D material.
\begin{figure}
\center
\includegraphics[width=.8\linewidth]{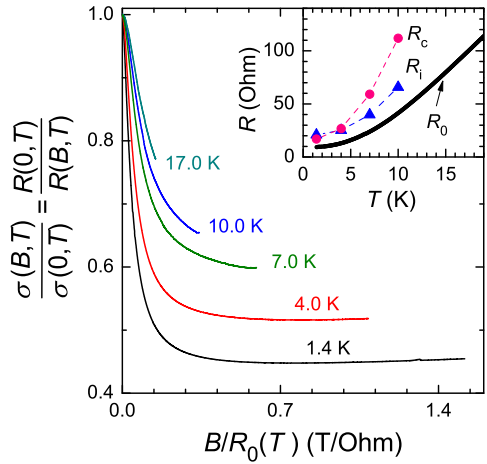}
\caption{(color online). Kohler plot of the normalized
interlayer conductivity of sample \# 1 for the field aligned
parallel to conducting layers obtained from field sweeps at
different temperatures. Inset: temperature dependence of the
zero-field resistance $R_0$ (thick line) and the resistances
of the coherent, $R_{\mathrm{c}}\propto 1/\sigma_{\mathrm{c}}$,
(circles) and incoherent,
$R_{\mathrm{i}}\propto 1/\sigma_{\mathrm{i}}$, (triangles) channels,
see text.}
\label{Kohler}
\end{figure}
According to Kohler's rule, the
magnetoresistance or, in our representation, magnetoconductivity
at different fields and temperatures should be just a function
of $B/R_0(T)$. This rule is strongly violated in Fig. 2:
the curves corresponding to different temperatures rapidly diverge
from each other, saturating at different levels \cite{comment-NMR}.
On the other hand, the saturation occurs at approximately the same
$B/R_0$, independent of $T$.

The described behavior suggests two parallel contributions in the
conductivity:
\begin{equation}
\sigma(\boldsymbol{B},\tau) =
\sigma_{\mathrm{c}}(\boldsymbol{B},\tau) +
\sigma_{\mathrm{i}}(B_{\perp},\tau).
\label{sigma}
\end{equation}
Here, the first term on the right-hand side is the coherent Boltzmann
conductivity depending on both the strength and orientation of a magnetic
field \cite{kart04a,comment-WI}. In a field parallel to layers it decreases
proportional
to $(\omega_c\tau)^{\alpha}$ with $1\leq\alpha\leq2$ and, at a high enough
field, the second term in Eq. (\ref{sigma}) becomes dominant. We  associate
the latter with incoherent interlayer charge transfer. In agreement with
previous observations \cite{kart06}, the incoherent conductivity is
insensitive to the inplane magnetic field, however, it does depend on the
field component $B_{\perp}$ perpendicular to layers. This is why the
resistance increases, as the field is tilted from the direction parallel
to layers [3 T and 15 T curves in Fig. 1(a)]. At high fields, the total
conductivity is dominated by $\sigma_{\mathrm{i}}$ in a large angular
interval around $\theta=\pm 90^{\circ}$, which leads to a scaling behavior of
magnetoresistance: $R(B,\theta)=R(B\cos\theta)$ \cite{kart06}.

Evaluating the relative contribution of $\sigma_{\mathrm{i}}$ to the total
zero-field conductivity $\sigma(0,T)$ of sample \# 1 from the level, at
which the curves in Fig. 2 come to saturation, and using the $R_0(T)$
data plotted in the inset of Fig. 2, one can extract separately the
temperature dependences of the classical Boltzmann (empty circles in the
inset) and incoherent (filled triangles) channels. Note that even the
anomalous, incoherent channel shows a metallic behavior.

The two-channel model provides a natural explanation for the anomalous
dip in the AMR found, at certain conditions, on very clean samples of
$\alpha$-(BEDT-TTF)$_2$KHg(SCN)$_4$.
In such samples, the conductivity is dominated by
$\sigma_{\mathrm{c}}$ as long as the inplane field component parallel
to the open Fermi sheets is small. This is, in particular, reflected
in the shape of the small-$\varphi$ AMR of sample \# 2 
shown in Fig. 1(b): the angular dependence in the vicinity of
$\theta = \pm 90 ^{\circ}$ is rather flat and shows a narrow
peak, revealing a coherent 3D Fermi surface \cite{kart04a}. When the 
inplane field
component is turned from the $a$ axis to the $c$ axis, which is parallel to
the open Fermi sheets, the coherent conductivity rapidly drops down
\cite{kart04a}. Under these conditions, the incoherent channel
$\sigma_{\mathrm{i}}$ may become important, which leads to the anomalous
dip structure in the $b^{\ast}c$-rotation patterns of the AMR even for
relatively clean samples \cite{kart06}. In the same way can be interpreted
the $90^{\circ}$ dips observed in the AMR of clean samples of other highly
anisotropic compounds, like (TMTSF)$_2$X with X=PF$_6$ \cite{kang07} and
ReO$_4$ \cite{kang03b} or
$\beta^{\prime\prime}$-(BEDT-TTF)$_2$SF$_5$CH$_2$CF$_2$SO$_3$
\cite{wosn02}.

The proposed model also explains the temperature dependence of
the crossover field $B_c$ observed in the experiment on (TMTSF)$_2$PF$_6$
\cite{chas98}. Indeed, while the contribution of the coherent channel
decreases with increasing temperature, as seen from Fig. 2, it still remains
significant at $T\sim 10$~K. At the same time, the scattering rate,
which is proportional to the resistance of the coherent channel, grows by
about an order of magnitude between 1.4 and 10~K (see inset in Fig. 2).
Therefore, a much higher field is necessary at 10~K for "freezing out"
$\sigma_{\mathrm{c}}$ and making the incoherent channel dominant in the
field and angular dependence of magnetoresistance.

Turning to a possible origin of the incoherent conduction channel,
its metallic behavior apparently comes into conflict with the existing
theories of incoherent interlayer charge transfer (see
\cite{lund03,ho05,maslov,abri99} and references therein) predicting an
insulating temperature dependence.
In addition, those theories do not account for the
significant dependence of $\sigma_{\mathrm{i}} $ on magnetic field normal
to layers. To comply with the experimental observations, we propose to
consider elementary events of incoherent interlayer hopping via
local centers, such as resonance impurities \cite{abri99,maslov},
in combination with diffusive intralayer transfer from one hopping
center to another. The essential requirement of our
model is that the volume concentration of hopping centers $n_{i}$ be
small, so that the average distance $l_{i}$ between them
along the 2D layers is much larger than the inplane mean free path
$l_{\tau }=v_{F}\tau $:
$
l_{i}=\left( n_{i}d\right) ^{-1/2}\gg l_{\tau }
$.
This condition, being opposite to the model \cite{maslov},
looks reasonable, since the concentration of resonant impurities (i.e. those
impurities which form an electron level with energy close to the Fermi
energy) is definitely much lower than the concentration of all kinds of
impurities. The current through each hopping center is limited by the
resistance $R_{\perp }$, which contains two in-series elements:
\begin{equation}
R_{\perp }=R_\mathrm{hc}+R_{\parallel }.
\label{R}
\end{equation}
The first part, $R_\mathrm{hc}$, is the hopping-center resistance itself,
which is almost independent of magnetic field and can have a weak
nonmetallic temperature dependence $R_\mathrm{hc}\left( T\right)$. The
second part, $R_{\parallel }$, is the intralayer resistance, which
comes out because the electrons must travel along the conducting
layer over a distance $\sim l_{i}$. In the limit $l_{i}\gg l_{\tau
}$, the 2D intralayer current density $\boldsymbol{j}\left(
\boldsymbol{r}\right) $ at each point is proportional to the electric field
$\boldsymbol{E}\left( \boldsymbol{r}\right) $ at this point:
$j_{\alpha }\left( \boldsymbol{r}\right) =
\sigma _{\alpha \beta }d\,E_{\beta }\left( \boldsymbol{r}\right)$,
where $\sigma _{\alpha \beta }$ is just the macroscopic 3D inplane
conductivity. For simplicity, we consider an isotropic inplane
conductivity:
$\sigma_{\alpha \beta }=\sigma _{\parallel }\delta _{\alpha \beta }$.

Since the charge density does not change with time, the inplane current
must satisfy $\text{div}\boldsymbol{j}\left( \boldsymbol{r}\right) =0$
everywhere except the hopping center spots. In the vicinity of each hopping center,
the current and electric field are roughly axially symmetric and given by
\begin{equation}
\boldsymbol{E}\left( \boldsymbol{r-r}_{i}\right) =
\frac{\boldsymbol{j}\left( \boldsymbol{r-r}_{i}\right) }
{\sigma _{\parallel }d}=\frac{I_{0}}{2\pi \sigma _{\parallel }d}
\frac{\left( \boldsymbol{r-r}_{i}\right) }
{\left\vert \boldsymbol{r-r}_{i}\right\vert ^{2}},
\label{2}
\end{equation}
where $I_{0}$ is the current through the hopping center located at point
$\boldsymbol{r}_{i}$. $R_{\parallel }$ is determined by the inplane mean voltage
drop between two successive hopping centers:
\begin{equation}
I_{0}R_{\parallel }\left( T\right) \simeq
2\int_{l_{\tau }}^{l_{i}}\boldsymbol{E}\left( r\right) dr=
\frac{I_{0}\ln \left( l_{i}/l_{\tau }\right) }
{\pi \sigma_{\parallel }d}.
\label{dV}
\end{equation}
As the lower cutoff in the integral (\ref{dV}) we take the mean free path
$l_{\tau }$, which neglects the resistance of the ballistic region
$\left\vert \boldsymbol{r-r}_{i}\right\vert <l_{\tau }$ around each
impurity. Since the ballistic conductivity is much higher than the
diffusive one, this approximation should work well, at least when
$\ln \left( l_{i}/l_{\tau }\right) \gtrsim
\ln \left( l_{\tau}/d\right) $.

The mean voltage drop between two adjacent
conducting layers is 
$E_{0}d=I_{0}\left( R_\mathrm{hc}+R_{\parallel}\right)$, 
where $E_{0}$ is the external electric field 
perpendicular to the layers. The total current density in the
interlayer direction is $j_{t}=I_{0}n_{i}d=\sigma _{\mathrm{i}}E_{0}$,
yielding the interlayer conductivity:
\begin{equation}
\sigma _{\mathrm{i}}=\frac{\pi \sigma _{\parallel }n_{i}d^{3}}{\pi
d\sigma _{\parallel }R_\mathrm{hc}+\ln \left( l_{i}/l_{\tau }\right) }.
\label{szz}
\end{equation}

The present simple model can be generalized by
including the distribution of the hopping centers
$n\left[ R_\mathrm{hc}\left( T\right) \right] $
and performing integration over $R_\mathrm{hc}\left( T\right) $.
The exact result will depend on the particular physical model of the
hopping centers. In the trivial case of short-circuiting the
layers (e.g., by dislocations),
$R_\mathrm{hc} \approx 1/d\sigma _{\parallel }$
and $\sigma _{\mathrm{i}}$ should be just proportional to the intralayer
conductivity. The fact that the temperature dependence of the incoherent
channel is considerably weaker than that of the coherent one
(inset in Fig. 2), implies that $R_\mathrm{hc}$ is larger than
$1/d\sigma _{\parallel }$ and only slightly varies with (or is
independent of) temperature. Such conditions can be fulfilled if the
hopping occurs via resonance impurities \cite{maslov}. Further,
both $R_\mathrm{hc}$ and $1/\sigma _{\parallel }$ are largely
insensitive to the inplane magnetic field in the incoherent regime
whereas the significant dependence on the out-of-plane field
component obviously comes from the intralayer conductivity. Thus,
the proposed mechanism is consistent with the main features of
the incoherent channel observed in the experiment.

In conclusion, we have shown that the anomalous behavior of the
angle-dependent interlayer magnetoresistance in the highly
anisotropic layered metal $\alpha$-(BEDT-TTF)$_2$KHg(SCN)$_4$
can be described by parallel contribution of two conduction
channels, $\sigma_{\mathrm{c}}$ and $\sigma_{\mathrm{i}}$,
providing, respectively, coherent and incoherent interlayer
charge transfer. A sufficiently high inplane component of
magnetic field changes the proportion of
$\sigma_{\mathrm{c}}$ and $\sigma_{\mathrm{i}}$ in favor
of the latter, thus causing an apparent dimensional
crossover. However, by contrast to the field-induced
confinement scenario \cite{clarke1}, this
crossover does not imply a change in the dynamic properties
of charge carriers. The proposed model is able
to explain not only the observed crossover but also anomalous
features found in other layered metals situated in the
transient region between the fully coherent and incoherent
transport regimes. 

The work was supported in part by DFG-RFBR Grant 
No. 436 RUS 113/926, by MK-4105-2007.2 and RFBR 06-02-16551.


\begin{thebibliography}{99}
\bibitem[{com({\natexlab{a}})}]{comment-NMR}
\bibinfo{note}{A careful examination reveals a slight, $\simeq 1$\%, parabolic
  increase of conductivity at high fields, for $T=1.4$ and 4~K. We will not
  consider this small effect here.}
\bibitem[{com({\natexlab{b}})}]{comment-WI}
\bibinfo{note}{This term also accounts for the so-called "weakly incoherent"
  regime \cite{kenz98} when the electron momentum is only defined within layers
  but is conserved in the elementary event of interlayer tunneling.}
\end{thebibliography}

\end{document}